\documentstyle{article}
\title{
\begin{flushright}
{\bf\normalsize   UB-ECM-PF 97/02}\\
\end{flushright}
\bf String tension in\\
gonihedric 3D Ising models
}

\author{
         {\it M. Baig}\\
	 IFAE, Universitat Aut\`onoma de Barcelona\\
	 Edifici C\\
	 08193 Bellaterra\\
	 Spain\\
	 \\
 {\it D. Espriu} \\
	 DECM and IFAE, Universitat de Barcelona\\
         Diagonal 647\\
         08028 Barcelona\\
         Spain\\
    \\
        {\it D.A. Johnston} \\
         and \\
	 {\it Ranasinghe P. K. C. Malmini $^{(a)}$}\\
         Dept. of Mathematics\\
         Heriot-Watt University\\
	 Riccarton,
         Edinburgh, EH14 4AS\\
         Scotland}
\begin{document}
  \maketitle
                      {\Large
                      \begin{abstract}
%
For the 3D gonihedric Ising models defined by Savvidy and Wegner
the bare string tension is zero and the energy of a spin interface
depends only on the number of bends and self-intersections,
in antithesis to the standard nearest-neighbour 3D Ising action.
When the parameter $\kappa$ weighting the self-intersections
is small the model has a first order transition
and when it is larger the transition
is continuous.
In this paper we investigate the scaling of the renormalized string tension,
which is entirely generated by fluctuations,
using Monte Carlo simulations
for $\kappa=0.0,0.1,0.5$ and $1.0$.
The scaling of the string tension allows us to obtain an estimate
for the critical exponents $\alpha$ and $\nu$
using both finite-size-scaling
and data collapse for the scaling function.
The behaviour of the string tension
when the self-avoidance parameter $\kappa$ is small
also clearly demonstrates
the first order nature of the transition in this case,
in contrast to
larger values.
Direct estimates of $\alpha$ are in good agreement with those
obtained from the scaling of the string tension. We have also measured
$\gamma/\nu$.
\\ \\ \\ \\ \\ \\
$(a)$ {\it Permanent Address:} \\
Department of Mathematics\\
University of Sri Jayewardenepura\\
Gangodawila, Sri Lanka.
%
                        \end{abstract} }
%
  \thispagestyle{empty}
%
%
  \newpage
%
                  \pagenumbering{arabic}

\section{Introduction}

The phase structure of Ising models with extended range
interactions in two and three dimensions is very rich
in general \cite{9,10}. In 3D when the spin
interfaces are regarded as describing a gas of closed surfaces
one gets a similarly rich diagram
for morphological transitions in an equivalent plaquette
surface model \cite{9a}. The relation between the weighting
of area, right-angled bends and intersections
in a plaquette surface model and the couplings of an Ising model
with nearest neighbour, next to nearest neighbour
and plaquette interactions on the dual lattice is
explicitly known \cite{9}. This equivalence between
an Ising model and a surface gas is particularly
convenient for performing Monte Carlo simulations,
where the Ising model transcription offers obvious
practical advantages.

In this paper we are interested in a class
of Ising model hamiltonians which assign
zero weight to the area of an interface.
Any surface tension that appears in the model
is thus generated by fluctuations. Higher spin
models with zero bare surface tension
\cite{Sch} have been investigated in some detail,
particularly in the context of wetting transitions
\cite{Die}, but the gonihedric Ising model
realization is still relatively unexplored.
The genesis of the gonihedric Ising models
lies in a random surface model developed by
Savvidy et al. \cite{1}
where the action for a triangulated 2D surface embedded in ${\bf R}^d$
is given by
\begin{equation}
S = {1 \over 2} \sum_{<ij>} | \vec X_i - \vec X_j | \theta (\alpha_{ij}),
\label{e4a}
\end{equation}
the sum being over the edges of some triangulated surface,
$\theta(\alpha_{ij}) = | \pi - \alpha_{ij} |^{\zeta}$,
$\zeta$ is some exponent (smaller than one, or else the model
is ill-defined \cite{jons}),
and $\alpha_{ij}$ is the dihedral angle between adjacent triangles.
This action was proposed to cure some of the diseases
of other triangulated random surface theories, which have
proved remarkably
reluctant to yield smooth continuum limits.

Taking the discretization a stage further and discretizing
the target space,
which thus becomes ${\bf Z}^d$, the authors of
\cite{7,8,8a,8b} rewrote the resulting theory
as precisely the sort of extended Ising model
considered in \cite{9,10}.
Specializing to three dimensions, the energy
of a plaquette surface in the gonihedric model is given by
by
$E=n_2 + 4 \kappa n_4$, where $n_2$ is the number
of links where two plaquettes meet at a right angle and
$n_4$ is the number of links where four plaquettes
meet at right angles.
$\kappa$ is a free parameter which determines the relative
weight of the intersections compared with the right-angled bends.
The results of \cite{9} show that an Ising Hamiltonian which
gives
$E=n_2 + 4 \kappa n_4$ contains nearest neighbour ($<i,j>$),
next to nearest neighbour ($<<i,j>>$) and round a plaquette ($[i,j,k,l]$)
terms
\begin{equation}
H= 2 \kappa \sum_{<ij>}^{ }\sigma_{i} \sigma_{j} -
\frac{\kappa}{2}\sum_{<<i,j>>}^{ }\sigma_{i} \sigma_{j}+
\frac{1-\kappa}{2}\sum_{[i,j,k,l]}^{ }\sigma_{i} \sigma_{j}\sigma_{k} \sigma_{l}.
\label{e1}
\end{equation}
Of course it should be pointed out that the equivalence
between eqs. (\ref{e4a}) and (\ref{e1}) is only at a very
intuitive level. Ising surfaces have a degree of self-avoidance,
which the gonihedric action has not.
The resolution of the resulting spin configurations in terms of
plaquette surfaces is ambiguous too (do the surfaces touch or cut
each other?), but for large enough (additional) self-avoidance coupling
$\kappa$ the
distinction may be irrelevant.
The ratio of coefficients that appear in eq.(\ref{e1}) is
rather particular, it corresponds to the so-called
disorder variety as calculated in the mean field approximation
\cite{dis}.

The
above  hamiltonian displays a ``flip'' symmetry
-- a plane of spins may be flipped at zero energy cost if it does not
intersect any other planes.
This symmetry poses some problems in Monte Carlo
simulations when one is attempting to
measure e.g. magnetic exponents as
lamellar low temperature configurations
with arbitrary interlayer spacing render the standard
magnetization
\begin{equation}
M = \left< {1 \over L^3} \sum_i \sigma_i \right>.
\label{ord}
\end{equation}
meaningless \footnote{Although a low temperature expansion
suggests that the lamellar state actually has a slightly  higher
energy than the purely ferromagnetic state \cite{beppe}, Monte-Carlo
simulations with periodic boundary conditions persistently
give lamellar configurations.}.
The solution adopted in \cite{desmal,desmal1} was to fix
three perpendicular planes of spins,
which provided a sufficient penalty to suppress the lamellar state
degeneracy
and still allowed one to retain the periodic boundary conditions
which minimize finite size effects.

The phase structure of the hamiltonian in eq.(\ref{e1}) has been explored
by both Monte Carlo \cite{desmal,desmal1} and cluster-variational (CVPAM) methods
\cite{beppe,beppe2} with similar results: there
is a single transition
from a paramagnetic high temperature phase to 
(with appropriate boundary conditions in the Monte Carlo case) 
a ferromagnetic phase.
The
transition point appears to be independent of $\kappa$ for $\kappa$
sufficiently large. The transition
is continuous for these values of $\kappa$. The critical indices
are also independent of $\kappa$. However,
in the vicinity
of $\kappa=0$, where the model displays
an additional ``antiferromagnetic''
symmetry, the transition becomes first order \cite{desmal1}.
A modified mean-field approach \cite{desmal,9}
also predicts the same phase structure, but
underestimates $\beta_c$ and gives
a much stronger variation of $\beta_c$ with $\kappa$
than is seen in the simulations and CVPAM approach.
It was suggested in \cite{beppe2} that consideration of a larger space
of coupling constants for the terms in eq.(\ref{e1})
indicated that the observed exponents were effective exponents
arising from the proximity of the transition
point to the critical end point
of a paramagnetic-ferromagnetic line. 

It was remarked in earlier work by Savvidy et.al. \cite{8}
that $\beta_c$ was close to that of the standard two-dimensional
Ising model with nearest neighbour interactions on a square lattice
and the simulations in \cite{desmal} found
$\gamma/\nu=1.79(4)$, close to the Onsager 2D Ising value
of 1.75. These authors also found
$\nu=0.8(1)$\footnote{$\nu$ is inadvertently transposed
with $1/\nu$ in \cite{desmal}.}, not far off the
2D Ising value $\nu=1$. However, the value of $\beta$ was estimated
to be a lot smaller than the Onsager value of 0.125.

Direct and finite size scaling
fits on the specific heat data in \cite{desmal} for $\alpha$
were not very reliable due to the presence of the unknown
analytic part in the specific heat, which leads to an extra adjustable
parameter in the fits. Part of the motivation
for the current work was to obtain an alternative
estimator for $\alpha$ that sidestepped these problems.
Another factor was the
apparent impossibility of finding discretized surface models
exhibiting a proper scaling of the string tension.
The ordinary 3D Ising model has a scaling string tension,
but it is unclear to which continuum
surface model it corresponds. It is thus
interesting to find equally simple models with different
scaling properties particularly if, as for the one presented
here, they have a nice geometrical interpretation.
An additional reason to undertake this project
was that, if present, the
string tension in the gonihedric models should be
entirely generated by fluctuations, so the simulations
allow one to confirm that the standard scaling
ansatze still apply in such a case.
Finally, the behaviour of the string tension
is an excellent indicator of a
first order transition, so a simulation
at, or around, $\kappa=0$ can confirm the first order
nature of the transition in this case. We have hypothesized
in \cite{desmal1} that the apparent impossibility of
defining a continuum limit for ``ghost" surfaces
($\kappa=0$ implies no self-avoidance) may be related
to the fact that eq. (\ref{e4a}) leads to problems if
$\zeta=1$, which the Ising discretization naively corresponds to.

We now move on to discuss the measurement of the surface tension
and extraction of estimates for $\alpha$ and $\nu$.

\section{Surface Tension}

Normally when one is trying to measure a string tension in an Ising
like model it is sufficient to use anti-periodic boundary conditions
in one direction
because this guarantees the presence of an interface.
The string tension can then
be defined by the difference between the bulk partition function $Z_0$
and the partition function with an interface $Z_I$
\begin{equation}
\sigma = { 1 \over L^2} \log \left( { Z_I \over Z_0 } \right)
\end{equation}
where we have assumed that we have a square interface spanning
an $L \times L$ boundary.
However, as we have seen, the gonihedric Ising models possess a symmetry
which allows planes of spins to be flipped at no energy cost, so
antiperiodic boundary conditions are insufficient
to force an interface. Something more coercive, in the
form of the fixed, or perhaps
more accurately ``mixed'',  plus and minus spin
boundary conditions shown in Fig.1a
is
necessary in order to make sure an interface exists.
The fixed spins on the faces make sure that any flipped
spin planes will be penalized by a boundary energy and thus
discouraged. While this has the disadvantage
of greater finite size effects than antiperiodic
boundaries one can also use
fixed boundary conditions in calculating the
``bulk'' contribution $Z_0$, but now with all spins
plus in order to eliminate the interface as shown in Fig1b.
The string tension is given by a ratio of free energies,
so there should be a degree of cancellation in the
finite size effects for $Z_0$ and $Z_I$.

In practice one cannot measure a 
partition function or free energy directly
in the simulations, so one
considers the derivative of the string tension
with respect to $\beta$, which gives the
(internal) energy difference between configurations
with and without an interface.
\begin{equation}
{\partial \sigma \over \partial \beta} =
\Delta E =  L_z  \left( \left< E_{+-} \right> - \left< E_{++} \right> \right)
\end{equation}
where we have denoted the mixed boundary
conditions that give the interface by $\{+-\}$
and the fixed boundary conditions by $\{++\}$
 and the $\langle E \rangle $'s are energy densities.
The volume of the system is $L_z\times L\times L$. If
$L_z=L$,
the standard finite size scaling behaviour for the specific heat
$C \sim \tilde C_0 + \tilde C_1 L^{\alpha / \nu}$ and the relation
$C = \partial E / \partial \beta$ mean that one would expect
\begin{equation}
E \simeq E_0 + E_1 L^{\alpha -1 \over \nu}
\end{equation}
where the constant $E_0$ has its origin
in the regular term in the specific heat and
would be expected to appear
for both sets of boundary conditions. A
measurement of the energy
with a single set of
boundary conditions
thus gains nothing over
specific heat measurements
as there is still a constant term $E_0$.
However, if
one considers both mixed and fixed
boundary conditions one would expect the {\it same}
regular part $E_0$ for both sets.
A measurement of the energy difference
therefore eliminates $E_0$ and a simple power law fit
to
\begin{equation}
\Delta E = L \left( \left< E_{++} \right> -
\left< E_{+-} \right> \right) \propto L^{1 + {\alpha -1 \over \nu} }
 =
 L^A
\end{equation}
determines the exponent $A$.
Similar methods have been used
to extract $\alpha$ for the Heisenberg model by
Holm and Janke \cite{wolf}.

In order to extract further estimates
from our simulations
we also
considered the standard scaling ansatz \cite{priv}
for the string
tension on an asymmetric lattices. Then
\begin{equation}
\sigma \sim {1 \over L^2} \Sigma
\left(  t L^{1 / \nu}, {L_z \over L} \right),
\end{equation}
where $t = |\beta - \beta_c|/ \beta_c$. This gives directly
\begin{equation}
\Delta E  \sim  L^{ - 2+1/\nu} \Sigma^\prime
 \left( t L^{1 / \nu},{L_z \over L} \right),
\label{de}
\end{equation}
so at the critical point
\begin{equation}
\Delta E  \sim  L^{- 2+1/\nu} \tilde \Sigma^\prime
\left( {L_z \over L} \right).
\end{equation}
For $\kappa=0.5$ we simulated lattices with
 various aspect ratios, $x = L_z / L$,
and adjusted the exponent $A$ in a plot of $\Delta E\ L^{-A}$
in order to attempt to obtain a smooth scaling function
 $\tilde \Sigma^\prime (x)$.
The ``best'' curve then gives an estimate
of $A = - 2+1/\nu$
\footnote{By hyperscaling in $D=3$, $- 2+1/\nu = 1 + (\alpha -1)/\nu$,
so the two definitions of $A$ we have given are equivalent.}.
The approach can also be used on symmetric lattices away from
the critical point as, referring to eq.(\ref{de})
when $L_z=L$,
we can see that $\Delta E \ L^{2 - 1 / \nu}$
plotted against $t L^{1 / \nu}$ should collapse
the data for various lattice sizes and temperatures to give a
smooth scaling function for the correct choice of $\nu$.

\section{Simulations}

\subsection{Energy and Specific Heat Exponents}

The simulations were carried out on symmetric lattices
of size $10^3,12^3,15^3,18^3,20^3,22^3$ and $25^3$
for $\kappa=1.0$. For $\kappa=0.5$
lattices with various aspect ratios
of sizes $10^2 \times 10$, $12^2 \times 10$, $14^2 \times 10$,
.... $20^2 \times 10$, $16^2\times 20$, $20^2\times 20$,
$24^2\times 20$, and $16^2\times 30$, $20^2\times 30$,
$24^2\times 30$, $30^2\times 30$ were used. Similar
sizes were considered for
$\kappa=0$ and $\kappa=0.1$.
We carried out from 50K up to 1000K measurement sweeps
after allowing
a suitable amount of time for thermalization,
depending on the value of $\beta$ and the lattice sizes.
We used the tried and tested code
from \cite{desmal,desmal1}.
which performs a
simple Metropolis update. It
is worth remarking that although the
gonihedric action contains competing interactions
it might be worthwhile formulating a cluster update
for the model as there is no frustration present.
The magnetization, energy, susceptibility, specific heat
and various cumulants
were are all measured in the standard fashion.
For each lattice size we simulated separately
with fixed and mixed boundary conditions in order to allow
us to measure $\Delta E$ from the combined results.

Taking, for example, the $\kappa=1.0$ results on symmetric lattices,
we find that the curve of $\Delta E$ displays
a maximum, which we use as our estimator
for the pseudo-critical point \footnote{This
proved more stable than using the maximum slope
- i.e. maximum of the specific heat, and would
in any case be expected to have the same scaling properties.}.
The values of $\Delta E / L$
at the appropriate pseudocritical points
are shown in Table.1.

\begin{center}
\begin{tabular}{|c|c|c|c|} \hline
$L$   &  12 & 15 & 18              \\[.05in]
\hline
$\Delta E / L$  & 0.0369(28) & 0.0307(20) & 0.0281(23)      \\[.05in]
\hline
$\beta_c(L)$ & 0.3938(4) & 0.4095(1) & 0.4124(1) \\[.05in]
\hline 
\end{tabular}
\end{center}

\begin{center}
\begin{tabular}{|c|c|c|c|} \hline
$L$   &  20 & 22 & 25             \\[.05in]
\hline   
$\Delta E / L$  &  0.0235(25) & 0.0244(21) & 0.0222(25)     \\[.05in]
\hline   
$\beta_c(L)$ & 0.4198(1) & 0.4204(1) & 0.4227(1)
\\[.05in]
\hline
\end{tabular}
\end{center}
\vspace{.1in}
\centerline{Table 1: $\Delta E / L $ for $\kappa=1.0$, along
with the estimated pseudocritical temperatures.}

\bigskip
\noindent
A fit to these values gives $\Delta E  \sim L^A \sim L^{0.3(1)}$
with a $\chi^2/d.o.f.$ of 0.3. This estimate gives
a much lower value for $\nu$ ($\nu=0.44(2)$) than that
in \cite{desmal} which was obtained by comparison of $\gamma/\nu$
and $\gamma$ (as well as the scaling of the
pseudocritical points).
A plot of the scaling function for
the energy in Fig.2 for $\nu=0.44$ gives quite a
good collapse of the data, providing further evidence
in support of this lower value.

We have only three symmetric lattice simulations to carry
out a simple finite size scaling fit for $\kappa=0.5$,
which give 0.4(1) for the exponent
in the fit to $\Delta E$ vs $L$. We do have, on the other
hand,
a wide range of asymmetric lattice sizes on which to try out
the asymmetric scaling program outlined in the section.2. In Fig.3
we show a plot of
$\Delta E\  L^{-A} =
\tilde \Sigma^\prime \left( {L_z/ L} \right)$
against ${L_z/ L}$. By inspection the smoothest data
corresponds to $A \sim 0.46(4)$ which is consistent
with the finite size scaling estimate of $0.4(1)$ for $A$
above. Other values of $A$ are also plotted to show
how the scaling curve quickly deteriorates once one moves
outside the acceptable region for the exponent.
The estimated value of $A$ for $\kappa=0.5$ is thus somewhat higher
than the value $0.3(1)$ obtained
when $\kappa=1.0$, but given the errors it is not
possible to discern whether the exponents are different.
We believe, in fact, that they are the same.
In any case, the values we obtain for $A$ are clearly
different from the 2D and 3D Ising model values
($A=0$ and $A \simeq -0.4$, respectively).

It is possible to confirm that direct fits to the specific heat
exponent are in agreement with the above method.
Given that fits to $C \sim C_0 + C_1 t^{-\alpha}$
and $C \sim \tilde C_0 + \tilde C_1 L^{\alpha / \nu}$
are rendered untrustworthy by the presence of the
analytical terms another possible approach
is simply to plot $C^{-1/\alpha}$ vs $\beta$ for
various choices of $\alpha$. Provided $C_0$ is not
too large
this would be expected
to collapse all the various lattice size data onto a
straight line. This approach turns out to
work remarkably well for both $\kappa=0.5$ and $\kappa=1.0$.
In Fig.4 we plot the results of the best data
collapse for $\kappa=0.5$ with $\alpha=0.7$. Via hyperscaling,
this is consistent with the estimates
of $\nu$ appearing from the string tension scaling measurements.
The results for $\kappa=1.0$ are 
shown in Fig.5. The best
fit is again obtained for $\alpha=0.7$ 
and the error bars can be estimated in both cases by looking
at how the straight line plot deteriorates
as the exponent is varied. 
As we have noted, standard fitting techniques
for $\alpha$ are not particularly convincing
with our data, but a fit to all
the $\beta<\beta_c$ data 
for $C \sim C_0 + C_1 t^{-\alpha}$ using
our best estimate of $\beta_c$ gives 
$\alpha=0.5(1)$, admittedly with
poor quality, for both $\kappa=0.5$ and $1.0$.
In addition the estimated value of $C_0$, while
not zero, is certainly small and provides
justification for the plots in Figs. 4,5.
In summary, we would estimate our best fit to be 
$\alpha=0.7(1)$, which
supports the result obtained from the scaling
of the string tension ($\nu \sim 0.44$ translates
to $\alpha \sim 0.7$ with hyperscaling).

It is also worth looking at the $\kappa=0.0$ and $\kappa=0.1$
results for
$\Delta E$ in order to see the signals of a first order
transition in the scaling of the string tension. For a first order transition
one has a finite string tension at the transition point
and throughout the ordered phase. A step function in the
string tension $\sigma$ translates into a
delta function (centred at $\beta_c$) in the measured
quantity $\Delta E$,
This is precisely what is observed in the simulations
both at $\kappa=0$ itself and at small values of $\kappa$
such as $0.1$ as can be seen very clearly in Fig.6.
Even at $\kappa=0.1$
the sharpness of the observed peak indicates that the
transition is still
strongly first order.

\subsection{Magnetic Exponents}

The speculations in \cite{desmal} regarding the similarity
with the 2D Ising model were largely based on
exponents coming from the susceptibility $\chi \sim L^{\gamma / \nu}$
and $\chi \sim t^{- \gamma}$, although it was noted that the
estimate for magnetic exponent
itself $M \sim L^{\beta / \nu}$, was much smaller than
the Onsager 2D Ising value. It is possible to play a similar
data collapse game with the susceptibility
measurements to that performed with the specific
heat in the previous section in order to extract an
estimate for $\gamma$
- one simply plots $\chi^{-1/ \gamma}$ against $\beta$ for various
$\gamma$ until the best straight line plot is obtained.
In principle this should work even better than for the specific heat
because of the absence of an analytic term in the scaling
expression for the susceptibility, but the fixed and mixed boundary
conditions  that are imposed in the interest of obtaining an interface
appear to do rather greater violence to the magnetic quantities
than to the energetic ones, so the finite size effects
are large. The fixed boundary conditions give
marginally better
scaling behaviour than the mixed boundary conditions, so
in what follows
we use these. However, even in this case 
as a consequence of the large finite size effects
it is difficult to arrive at a precise direct determination
of $\gamma$. All we can say is that it is in the region
comprised between $\gamma=1$ and $\gamma=2$, which, admittedly,
is not saying much.
It is interesting that the CVPAM estimate for $\gamma$ (1.4)
which is unaffected by boundary condition and finite size considerations
is also relatively small, although the systematic errors
are difficult to assess in this case.

Nonetheless, finite-size scaling allows a much better
measurement of
 $\gamma/\nu$
for both $\kappa=0.5$ and $\kappa=1.0$.
One obtains in both cases
$\gamma/ \nu =2.1(1)$. The data used in the $\kappa=1.0$
fit on symmetric lattices is shown in Table.2 below, which produced
a fit of $2.1(1)$ with a $\chi^2 / d.o.f.$ of $0.18$.

\begin{center}
\begin{tabular}{|c|c|c|c|} \hline
$L$   &  12 & 15 & 18              \\[.05in]
\hline
$\chi_{max}$  & 9.4(3) & 14.9(6) & 22.7(1.8)     \\[.05in]
\hline
$\beta_c(L)$ & 0.3784(3) & 0.3919(4) & 0.3975(2)  
\\[.05in]
\hline
\end{tabular}
\end{center}

\begin{center}
\begin{tabular}{|c|c|c|c|} \hline
$L$   &  20 & 22 & 25             \\[.05in]
\hline
$\chi_{max}$  &  28.8(2.2) & 32.9(2.0) &  43.0(3.7)    \\[.05in]
\hline
$\beta_c(L)$ & 0.4027(3) & 0.4148(1) & 0.4144(1)
\\[.05in]
\hline
\end{tabular}
\end{center}
\vspace{.1in}
\centerline{Table.2: Data used in fitting $\gamma / \nu$ for $\kappa=1.0$.}

\bigskip
\noindent
For
 $\kappa=0.5$ on asymmetric lattices we perform a finite-size scaling analysis
similar to the one employed in determining the string tension scaling.
We search for a universal scaling function $X(L_z/L)$ by adjusting
the value of $\gamma/\nu$ in a 
plot of $X(L_z/L) \sim \chi L^{-\gamma/\nu}$.  The smoothest function is
also obtained when $\gamma/\nu=2.05$, in excellent agreement with
the $\kappa=1$ result. The function $X$ is plotted in
Fig.7. To show the sensitivity of the method, we also plot
$X$ for $\gamma/\nu=1.80$ which gives a markedly less
smooth curve.

A final word of warning:
while the estimates obtained are rather higher than the 1.79(4)
in \cite{desmal} for $\gamma / \nu$ at $\kappa=1.0$
it should be remembered
that a different set of boundary conditions (periodic,
with fixed {\it internal}
spin planes) was used there, and that these might be
expected to cause less
severe finite size effects for magnetic quantities.
The fixed boundary conditions do
even more violence to the magnetization itself
and we could obtain no reliable
estimates for either the magnetization
exponent $\beta$ or $\beta / \nu$
from finite size scaling, in contrast to
the periodic boundaries and fixed internal spin planes in \cite{desmal}.

\section{Discussion}

As we have already noted, from the results
it is clear that the 2D and 3D Ising
values for $A$ of $0$ and $\sim -0.4$
are
definitively excluded. This shows that the hypothesis floated
in \cite{desmal} that the critical behaviour of the model might
be related to that of the standard 2D Ising model is not
supported by the numerical evidence. There is good agreement
between the various methods we have used to extract the energetic
exponents ($\alpha$, $\nu$). The data collapse on both symmetric
and asymmetric lattices suggests that $\alpha =0.7(1)$ 
for both $\kappa=0.5$ and $1.0$ and (via both
hyperscaling and directly from the data collapse) $\nu$ is in
the region of $0.5$. We find a value around $2$ for $\gamma / \nu$
from the finite size scaling of the susceptibility, which implies
a smaller value of $\gamma$ ($\sim 1$)
than that estimated in \cite{desmal}.
As we have noted,
the estimate for $\gamma$ from the CVPAM
calculations \cite{beppe2} is similarly small.
The set of exponents we have arrived
are consistent with
$\alpha + 2 \beta + \gamma =2$. In addition hyperscaling,
$\alpha = 2 - \nu D$, appears to
be happily satisfied by the measured values of $\alpha$ and $\nu$.
The first order nature of the transition
for small $\kappa$ is also manifest in the sharp
spike in the measurements of
$\Delta E$ at $\beta_c$ for $\kappa=0.0,0.1$.

It is interesting that a similar value of
$\alpha=0.5$ appears in the vicinity
of the disorder variety for the
anisotropic triangular lattice Ising model \cite{hil,ghlm}.
The Bl\"ote and Hilhorst transcription
of the anisotropic triangular lattice Ising model as a
diamond covering of the lattice is particularly intriguing from
our viewpoint. In this formulation the
model can be viewed as a perspective view
from the $(1,1,1)$ direction of
a cubic lattice SOS model.
The excitations in the model appear as steps,
or strings of flipped diamonds
in the perspective view of an otherwise flat surface.
It is noteworthy that the weights which appear
in their model
are identical to weights for a restricted class
of surface configurations
in the gonihedric Ising model, namely: steps of
height one; no right angled
bends; no overhangs; and $\Delta_2$ (one of their parameters) zero.
Similar $\alpha=0.5$ singularities appear in related dimer
problems \cite{wu} which can be formulated as hexagonal lattice
vertex models.
Indeed,
it is possible to write down a vertex model that represents the
full gonihedric set of surface configurations as
viewed from a $(1,1,1)$ perspective but this does not
appear to be soluble.

Haldane and Villain \cite{hv} have pointed out that
a square root singularity would be expected to appear
generically
in systems with string-like, or striped, excitations
in two dimensions.
Given the values of $\alpha$ that we have estimated
from the string tension scaling here, it is not
inconceivable that we have a three dimensional
realization of these ideas in the gonihedric Ising
model.

\section{Acknowledgements}

R.P.K.C. Malmini was supported by Commonwealth Scholarship
SR0014. DE and DAJ were partially supported by EC HCM network
grant ERB-CHRX-CT930343. DE acknowledges the financial support of CICYT and
CIRIT through grants AEN95-0590 and GRQ93-1047,
respectively. 
MB acknowledges the financial support 
of CESCA and  the financial support of CICYT
through grant AEN95-0882.
DAJ would like to thank Wolfhard Janke for clarifying
discussions on the scaling form of $\Delta E$. DE would like to thank the
hospitality of the CERN TH Division where this work was finished.

\vfill
\eject

\clearpage \newpage
\begin{figure}[htb]
\vskip 20.0truecm
\includegraphics{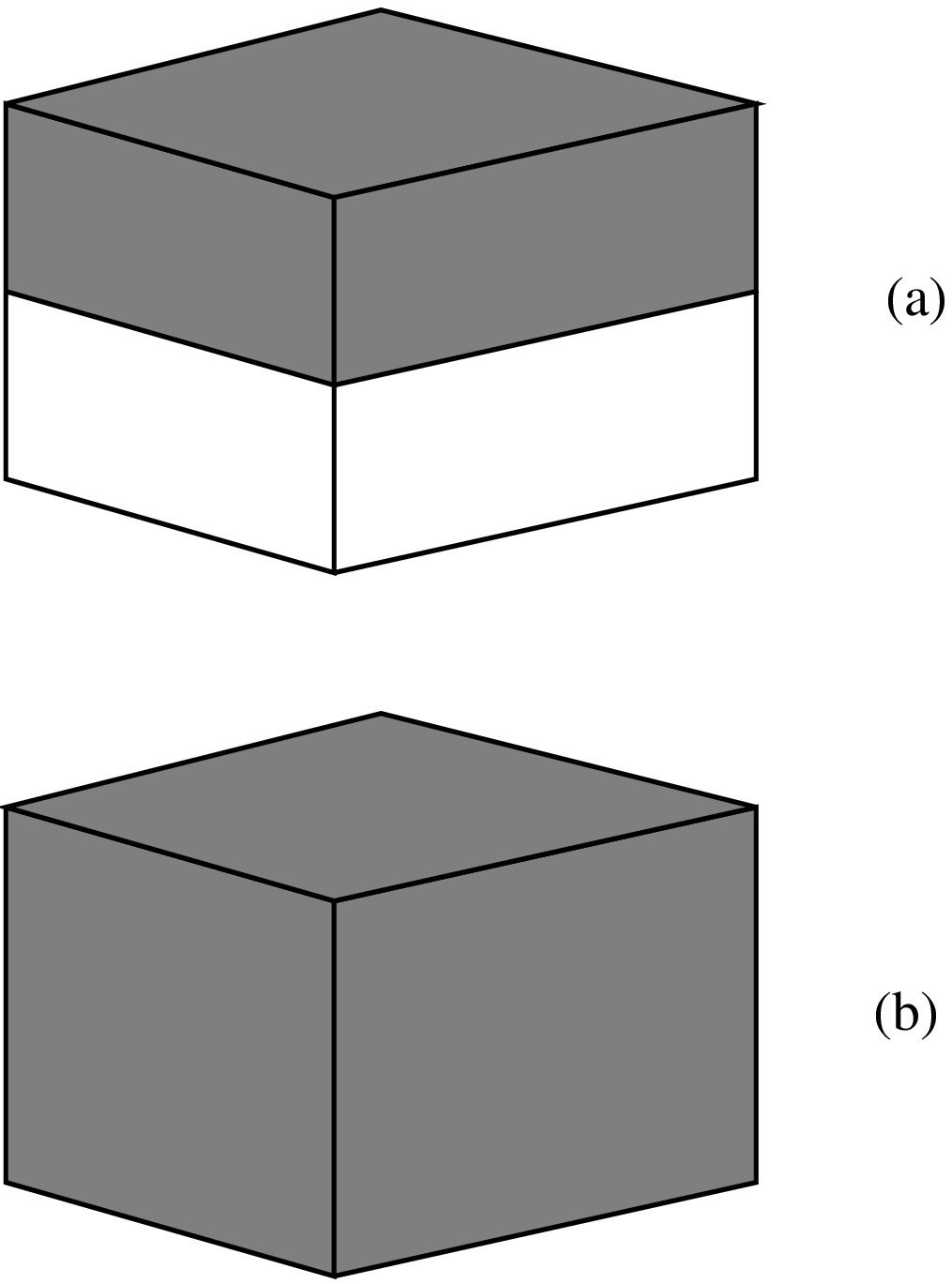}
\caption[]{\label{fig0} The boundary conditions
used in measuring the string tension, (a) is used
to produce an interface, (b) to ensure
a ferromagnetic phase. On the shaded surfaces the spins
are fixed to plus and on the white surfaces to minus.}
\end{figure}
\clearpage \newpage
\begin{figure}[htb]
\vskip 20.0truecm
\includegraphics{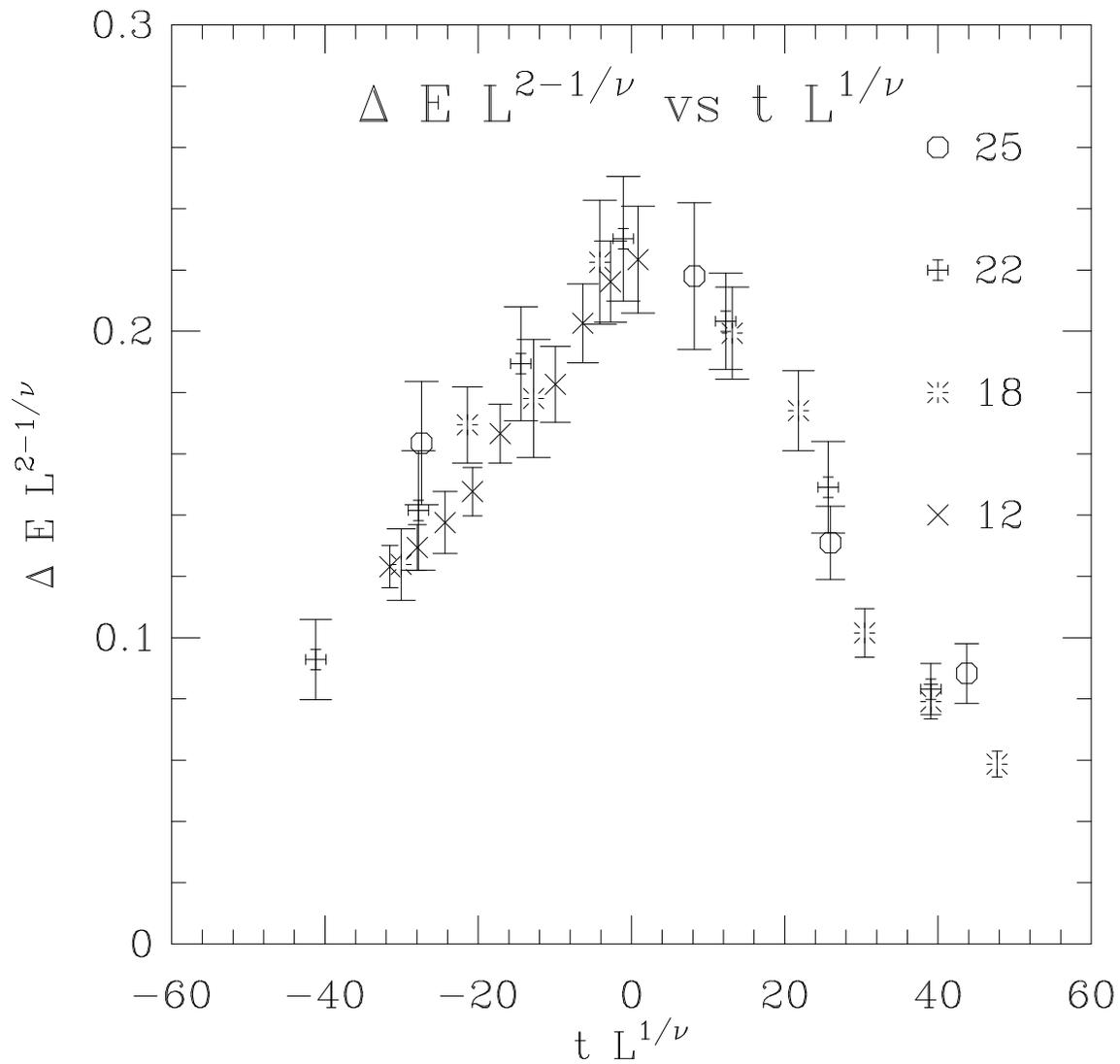}
\caption[]{\label{fig1} The scaling function on symmetric lattices
at $\kappa=1$, showing the good data collapse when $\nu=0.44$.
Some of the intermediate lattice sizes have been dropped
for clarity.}
\end{figure}
\clearpage \newpage
\begin{figure}[htb]
\vskip 20.0truecm
\includegraphics{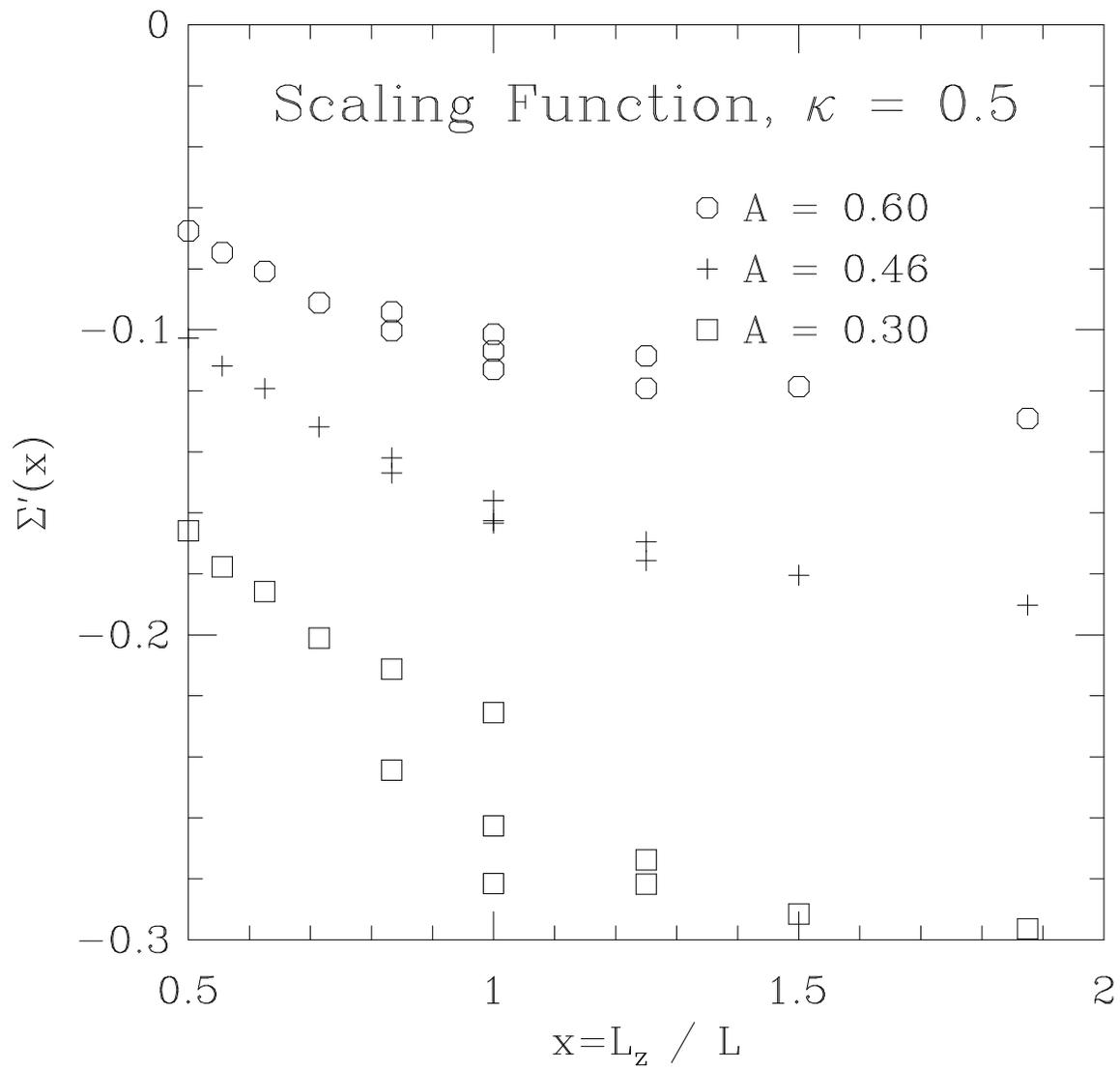}
\caption[]{\label{fig2} The asymmetric lattice scaling function
for $\kappa=0.5$ is
plotted for various choices of the exponent $A$ defined in the text.}
\end{figure}
\clearpage \newpage
\begin{figure}[htb]
\vskip 20.0truecm
\includegraphics{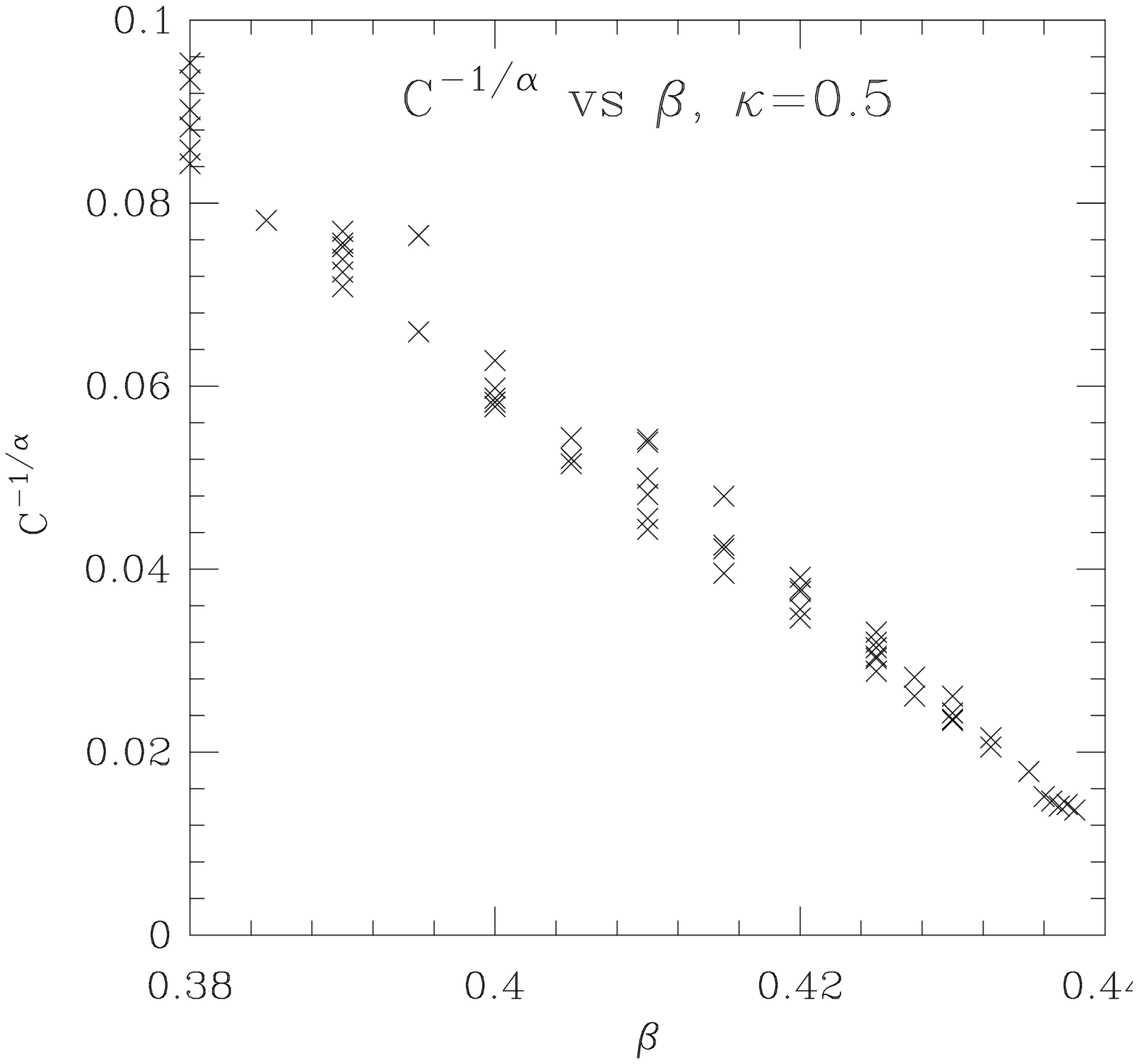}
\caption[]{\label{fig3a} $C^{-1 / \alpha}$ vs $\beta$ for the
best choice of $\alpha=0.7$ at $\kappa=0.5$}
\end{figure}
\clearpage \newpage
\begin{figure}[htb]
\vskip 20.0truecm
\includegraphics{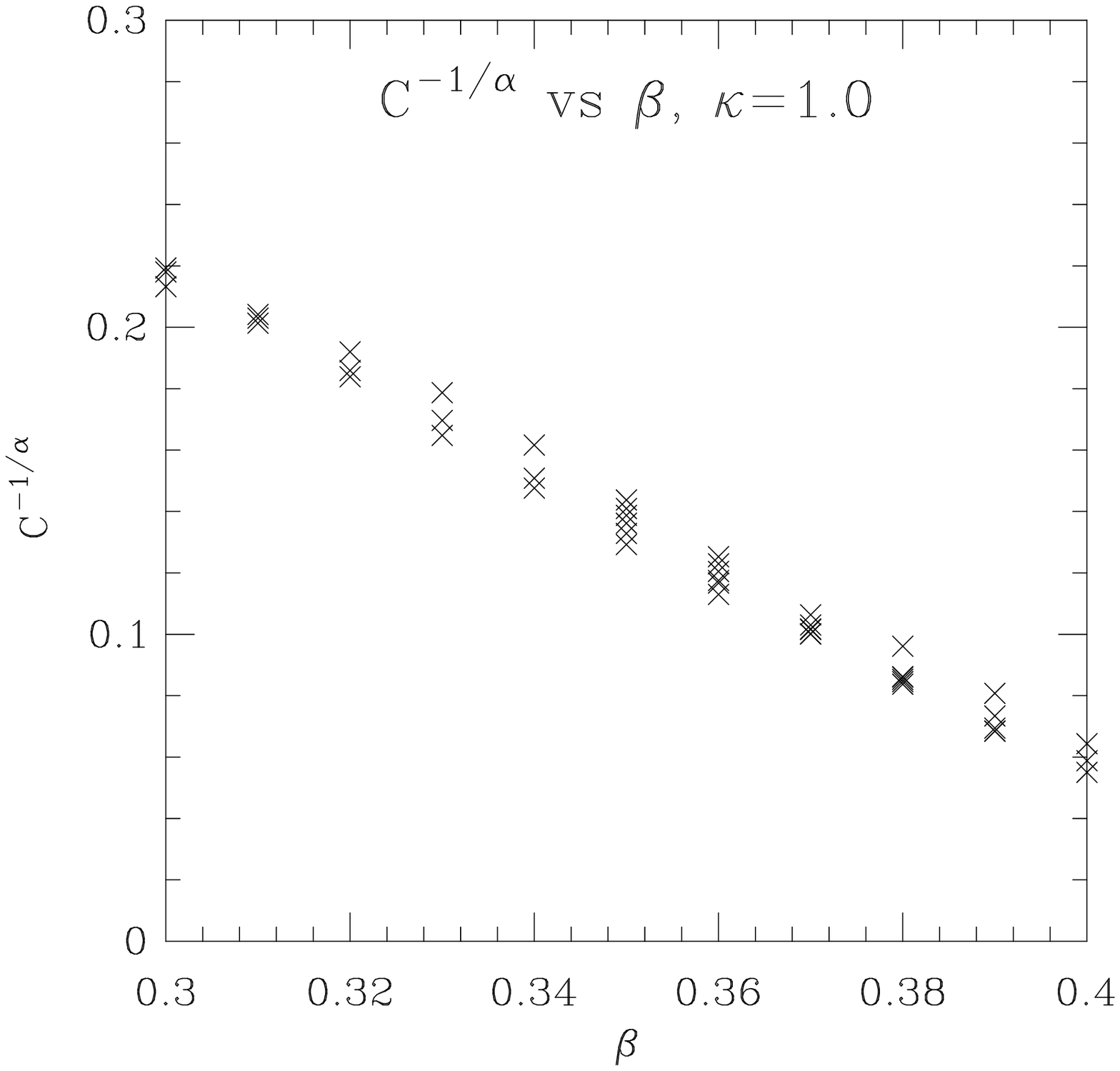}
\caption[]{\label{fig3b} $C^{-1 / \alpha}$ vs $\beta$ for the
best choice of $\alpha=0.7$ at $\kappa=1.0$}
\end{figure}
\clearpage \newpage
\begin{figure}[htb]
\vskip 20.0truecm
\includegraphics{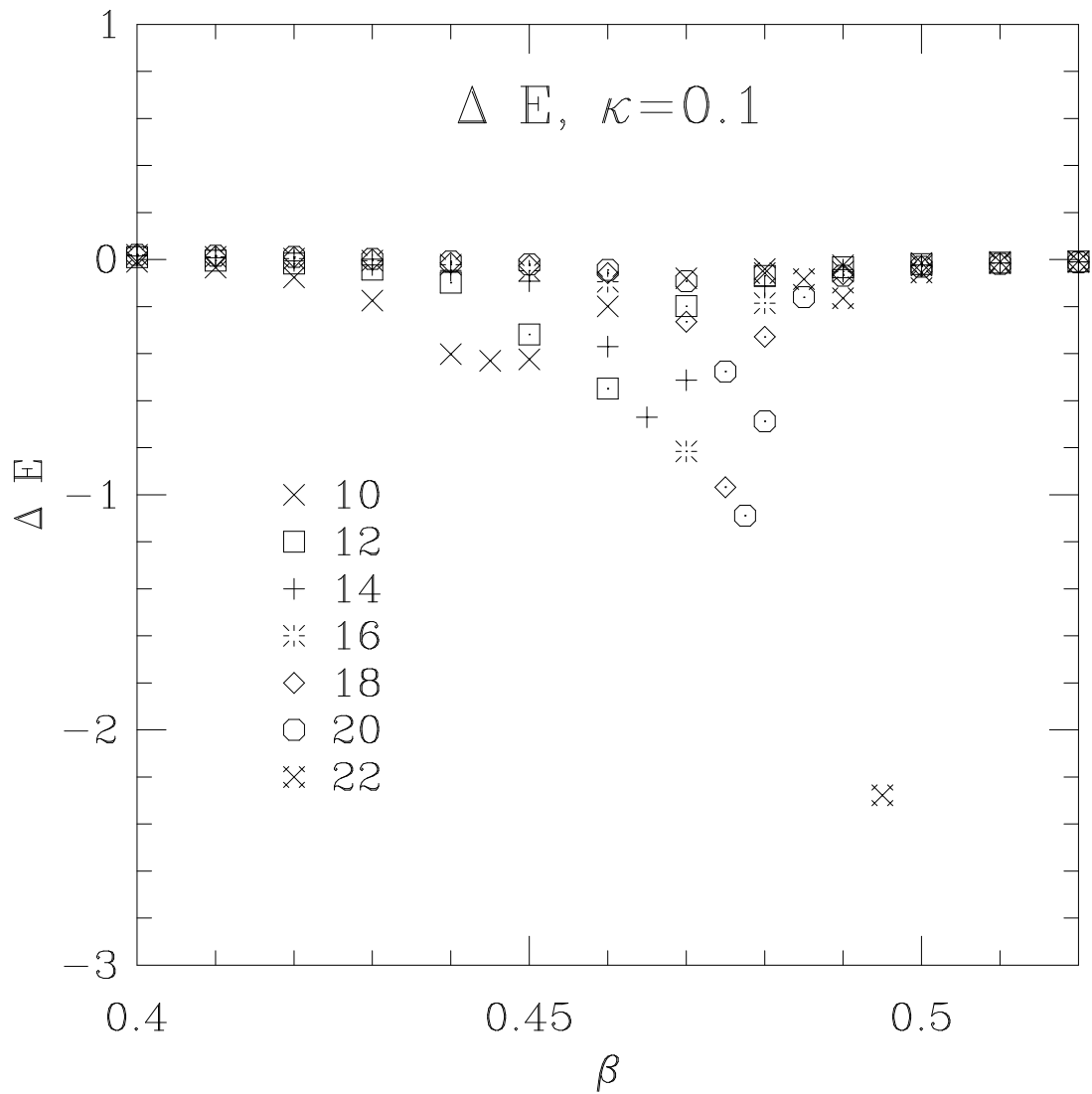}
\caption[]{\label{fig4} $\Delta E$ for $\kappa=0.1$
on various asymmetric $L^2 \times 10$ lattices. The sharp
peak is a clear indication of first order behaviour.}
\end{figure}
\clearpage \newpage
\begin{figure}[htb]
\vskip 20.0truecm
\includegraphics{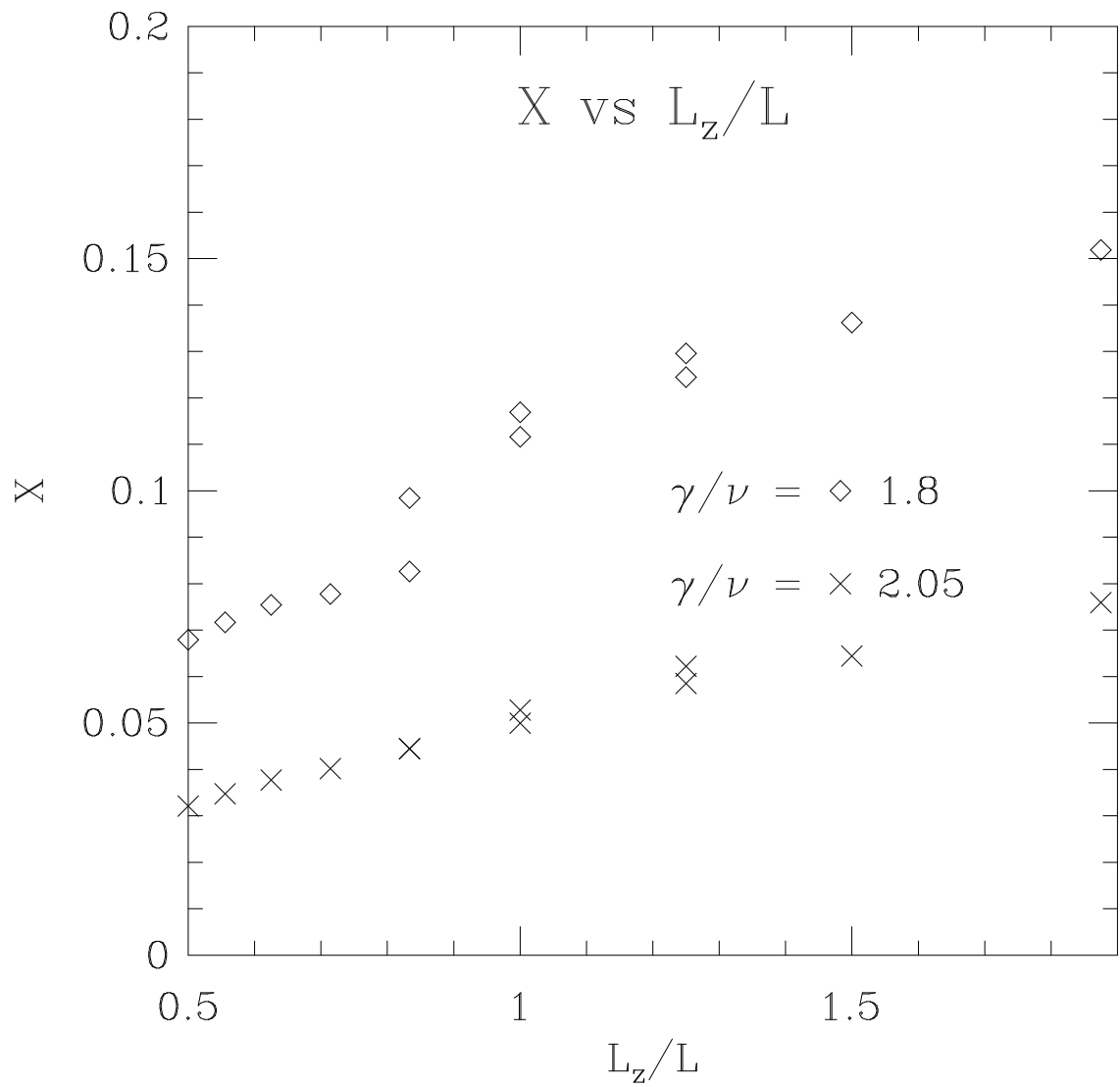}
\caption[]{\label{fig5} $X(L_z/L)$ for 
$\gamma/\nu=1.8,2.05$ at $\kappa=0.5$}
\end{figure}

\end{document}